\documentclass[aps,pra,reprint,amsmath,amssymb,superscriptaddress,nobibnotes, longbibliography]{revtex4-1}
\usepackage{graphicx,SIunits}
\usepackage{bm}     
\usepackage{hyperref} 
\usepackage{dsfont}    
\usepackage{color}

\usepackage{amsmath}
\usepackage{amsthm}
\usepackage{amssymb}
\usepackage{tikz}

\usetikzlibrary{decorations.pathreplacing}
\usepackage{braket}
\usepackage{dsfont}

\usepackage{color}

\begin{document}
	
	\title{Persistent quantum advantage with definite photon-number states \\ in lossy multiple-phase estimation}
	\author{Min Namkung}
	\affiliation{Center for Quantum Technology, Korea Institute of Science and Technology (KIST), Seoul, 02792, Korea}
	
	\author{Dong-Hyun Kim}
	\affiliation{Center for Quantum Technology, Korea Institute of Science and Technology (KIST), Seoul, 02792, Korea}
	\affiliation{Department of Physics, Yonsei University, Seoul 03722, Korea}
	
	\author{Seongjin Hong}
	\affiliation{Department of Physics, Chung-Ang University, Seoul 06974, Korea}

    \author{Changhyoup Lee}
    \email{changhyoup.lee@gmail.com}
    \affiliation{Korea Research Institute of Standards and Science, Daejeon 34113, Korea}
	
	\author{Hyang-Tag Lim}
	\email{hyangtag.lim@kist.re.kr}
	\affiliation{Center for Quantum Technology, Korea Institute of Science and Technology (KIST), Seoul, 02792, Korea}
	\affiliation{Division of Quantum Information, KIST School, Korea University of Science and Technology, Seoul 02792, Korea}

\date{\today} 
	
\begin{abstract}
Multiple-phase estimation exploiting quantum states has broad applications in novel sensing and imaging technologies. However, the unavoidable presence of a lossy environment in practical settings often diminishes the precision of phase estimations. To address this challenge, we propose an optimal multiple-phase estimation scheme that is inherently robust against photon loss, ensuring a persistent quantum advantage across all levels of photon loss. The scheme employs a multi-mode definite photon-number (DPN) state with weights optimized for given levels of photon loss. We theoretically demonstrate that the DPN state can sustain quantum enhancement in estimation precision under all levels of photon loss, compared to the classical benchmark that employs a coherent state input. The proposed scheme using DPN states generalizes earlier studies employing NOON states, which are only optimal when photon loss is small. We believe that our study, demonstrating persistent robustness to photon loss, paves the way for significant advancements in quantum-enhanced sensing technologies, enabling practical applications and quantum advantages in real-world scenarios. 
\end{abstract}
	
	
\maketitle

\section*{Introduction}
Quantum states offer the capability to estimate an unknown phase (or multiple phases) with precision surpassing the standard quantum limit (SQL) regarded as a classical limit. It has been shown that particular types of  discrete variable quantum states can achieve the Heisenberg limit~\cite{e.polino,m.szczy,j.f.haase,l.hwang,p.c.humphreys,l.zhang,l.zhang2,j.urrehman,m.namkung}, which is the fundamental precision limit allowed in quantum theory. Beyond theoretical formulations, not only single-phase but also various multiple-phase estimation schemes have been experimentally implemented with proper quantum states \cite{g.y.xiang,m.a.ciampini,z.-e.su,e.roccia,e.polino2,s.-r.zhao,s.hong,s.hong2}. Optimization of quantum states is thus crucial in quantum sensing and metrology to achieve quantum enhancement beyond the classical limit. 

The optimization of quantum states has so far been usually made in the absence of loss or noise in many cases. However, quantum states employed for multiple-phase estimation are hardly isolated from lossy environment, making their associated quantum enhancement be vulnerable to photon loss  \cite{j.-d.yue,r.demkowicz0,u.doner,r.demkowicz2}. For example, multi-mode NOON states achieve the Heisenberg limit in scenarios without losses~\cite{l.hwang,p.c.humphreys}, but their estimation precision cannot consistently surpass the SQL when photon loss occurs \cite{u.doner}. To mitigate the impact of photon loss, a multiple-phase estimation scheme has been proposed using multi-mode NOON states, optimized to maximize estimation precision~\cite{m.namkung}. Unlike previous approaches that rely on error mitigation \cite{k.yamamoto,b.koczor,w.j.huggins,h.kwon} or postselection \cite{drm}, the proposed scheme in Ref.~\cite{m.namkung} is able to reduce the effect of loss by optimizing a probe state without requiring additional quantum resources. Unfortunately, as verified in Ref.~\cite{m.namkung}, quantum enhancement of the NOON states is maximized for given levels of photon loss, but eventually diminishes as photon loss becomes significant, i.e., the proposed scheme is useful only when loss is moderate. This limitation is due to the nature of multi-mode NOON states, which involve only $N$ photons in each mode. This motivates the use of more general probe states such as definite photon-number (DPN) states which distribute $N$ photons across all modes~\cite{u.doner,r.demkowicz2}. DPN state generalizes NOON states, and its minimal form, i.e., the two-mode two-photon DPN state, has been experimentally implemented~\cite{m.kacprowicz}. Additionally, the approach proposed in Ref.~\cite{m.namkung} can be extended further, potentially yielding additional precision enhancements. This leads to the fundamental question: \textit{Can a scheme employing DPN states consistently outperform classical approaches across all levels of photon loss?}

In this work, we address the above question by studying a scheme that utilizes optimal multi-mode DPN states in lossy multiple-phase estimation. The $m$-mode $N$-photon DPN state is defined here as distributing $N$ photons across all modes. For the purpose, we employ the standard framework of quantum parameter estimation formulated in terms of the quantum Fisher information matrix (QFIM), which sets the asymptotic lower bound for estimation uncertainty. We demonstrate that, through optimization of the weights, DPN states consistently provide a quantum advantage over the classical limit in simultaneous multiple-phase estimation scenarios with photon loss. Furthermore, we elaborate on the behavior of the proposed scheme, showing that the optimal DPN states approximate NOON states proposed in Ref.~\cite{m.namkung} under moderate photon loss rates. We also examine the impact of photon loss occurring in individual modes, demonstrating that additional energy is required on the mode where the loss occurs. This suggests that, generally, the average photon numbers should be rebalanced according to the specific photon loss rates when optimizing DPN states. We believe that the results presented here can contribute to development of advanced noise suppression techniques applicable to novel quantum sensing applications.

\section*{Results}

\begin{figure}[t]
\centerline{\includegraphics[width=9cm]{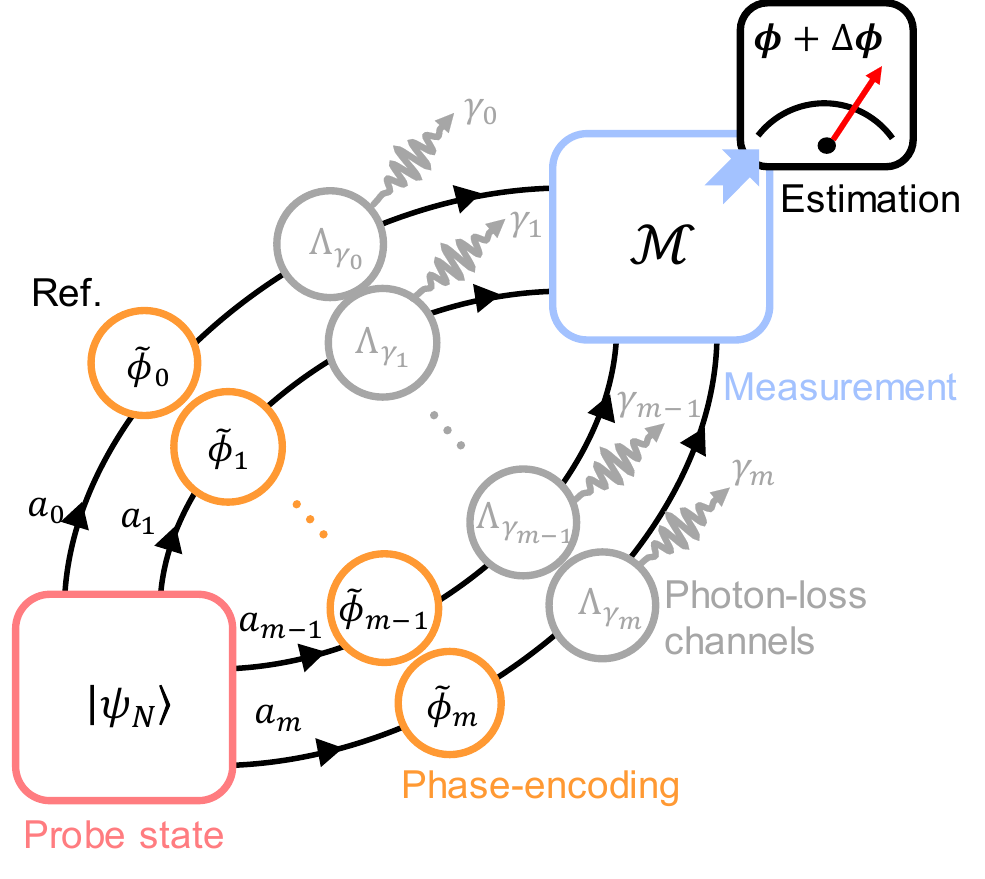}}
\caption{\textbf{Multiple-phase estimation scheme for estimating $m$ unknown relative phases $\bm{\phi}=\{\phi_1,\cdots,\phi_m\}$ with $\bm{\phi_j=\widetilde{\phi}_j-\widetilde{\phi}_0}$.} Here, it is assumed that a photon loss channel $\Lambda_{\gamma_j}$ with a photon loss rate $\gamma_j$ is on mode $j$. Each phase $\widetilde{\phi}_j$ is encoded in an $N$-photon probe state $|\psi_N\rangle$, and corresponding information is extracted therefrom by a measurement ($\mathcal{M}$). Then, each relative phase is estimated using an estimator based on measurement outcomes.}
\centering
\label{fig1}
\end{figure}

Let us begin with reviewing the theoretical background of multiple-phase estimation \cite{s.hong,s.hong2,j.urrehman,m.namkung}. The multiple-phase estimation scheme we consider is illustrated in Fig.~\ref{fig1}, where mode 0 is assumed to be a common reference mode and all the modes are subject to photon loss. A prepared probe state $|\psi_N\rangle$ consisting of $N$ photons passes through phase shifts described by a unitary operator $\hat{U}_{\rm \bm{\widetilde{\phi}}}=\bigotimes_{j=0}^{m}e^{i\widetilde{\phi}_j\hat{n}_j}$ with an unknown phase $\widetilde{\phi}_j\in\bm{\widetilde{\phi}}=\{\widetilde{\phi}_0,\widetilde{\phi}_1,\cdots,\widetilde{\phi}_m\}$ and the number operator $\hat{n}_j$, each defined on mode $j$. This transforms the probe state $|\psi_N\rangle$ into a phase-encoded state $|\psi_{N,\bm{\widetilde{\phi}}}\rangle=\hat{U}_{\rm \bm{\widetilde{\phi}}}|\psi_N\rangle$, followed by photon loss channels $\mathbf{\Lambda}=\bigotimes_{j=0}^{m}\Lambda_{\gamma_j}$ \cite{m.grassl}. Here, a photon loss channel $\Lambda_{\gamma_j}$ on mode $j$ with a photon loss rate $\gamma_j\in[0,1]$ is described as a fictitious beam-splitter, represented by mode transformation $\hat{a}_j\rightarrow\sqrt{1-\gamma_j}\hat{a}_j+\sqrt{\gamma_j}\hat{e}_j$ \cite{p.radmore} with a mode operator $\hat{a}_j$($\hat{e}_j$) on the signal (environment) mode $j$. Then, information about the encoded phases is extracted from the lossy phase-encoded probe state $\hat{\rho}_{N,\bm{\widetilde{\phi}}}=\mathbf{\Lambda}\left[|\psi_{N,\bm{\widetilde{\phi}}}\rangle\langle\psi_{N,\bm{\widetilde{\phi}}}|\right]$ through a measurement, followed by an estimation applied to the measurement outcomes.

The purpose of our scheme is to simultaneously estimate $m$-relative phases $\phi_j=\widetilde{\phi}_j-\widetilde{\phi}_0$, all with respect to the reference phase $\phi_0$. Note that the same estimation scenario has been studied in several previous works, but only in a lossless situation \cite{p.c.humphreys,j.urrehman,s.hong,s.hong2}. Considering such an estimation scenario, let us use $\bm{\phi}=\{\phi_1,\cdots,\phi_m\}$ as the parameters of interest to be estimated, i.e.,  $\tilde{\boldsymbol{\phi}}\rightarrow \boldsymbol{\phi}$ in the subscripts of the above expressions. The precision of simultaneous multiple-phase estimation can be evaluated in terms of total uncertainty $|\Delta\bm{\phi}|^2=\sum_{j=1}^{m}(\Delta\phi_j)^2$. If an unbiased estimator is considered, then the total uncertainty satisfies the following quantum Cramer-Rao inequality \cite{p.c.humphreys,s.hong,s.hong2,j.urrehman,j.liu2,l.pezze}:
\begin{align}\label{bd}
    |\Delta{\bm{\phi}}|^2\ge\frac{1}{\mu}\mathrm{Tr}\left[\mathbf{\bm{F}}_{\rm Q}^{-1}(\hat{\rho}_{\mathit{N},\bm{\phi}})\right],
\end{align}
where $\mu$ is the number of measurement repetition and $\mathbf{\bm{F}}_{\rm Q}(\hat{\rho}_{\mathit{N},\bm{\phi}})$ is QFIM. The quantum Cramer-Rao bound (QCRB) per a single measurement try is defined as $\mathrm{Tr}\left[\mathbf{\bm{F}}_{\rm Q}^{-1}(\hat{\rho}_{\mathit{N},\bm{\phi}})\right]$. If $\hat{\rho}_{\mathit{N},\bm{\phi}}$ contains zero eigenvalues, it is sufficient to construct the QFIM $\mathbf{\bm{F}}_{\rm Q}(\hat{\rho}_{\mathit{N},\bm{\phi}})$ on the support of $\hat{\rho}_{\mathit{N},\bm{\phi}}$ \cite{c.oh}. Note that, along with maximum likelihood estimation \cite{l.pezze,c.w.helstrom}, there always exists an optimal measurement that achieves the QCRB in the estimation of separable multiple phases. It means that the QCRB can be used to quantify the estimation precision achievable in quantum metrology.


\begin{figure*}[t]
\centerline{\includegraphics[width=18cm]{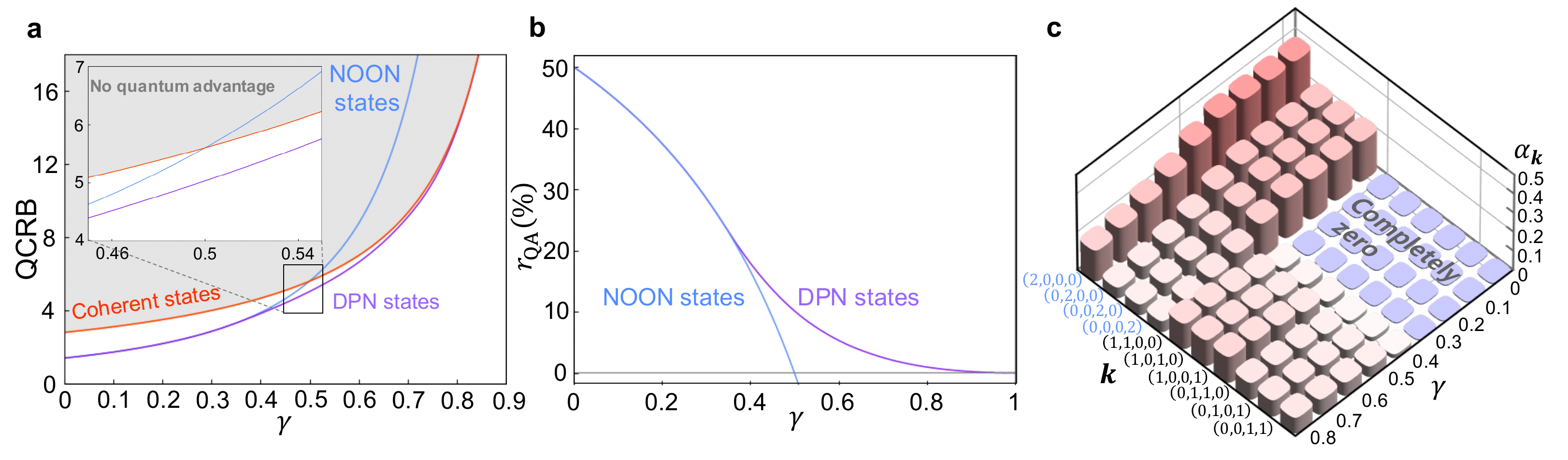}}
\caption{\textbf{Quantum advantage in estimating three relative phases.} \textbf{a} QCRB in estimating three phases with optimal four-mode two-photon probe states. Here, all photon loss rates are assumed to be equal to $\gamma$. \textbf{b} Quantum advantage of NOON and DPN states quantified by $r_{\rm QA}$ in Eq.~(\ref{QA}). \textbf{c} Optimal weights of four-mode DPN states studied in panel \textbf{a}.}
\centering
\label{fig2}
\end{figure*}

The probe state considered in this work is a multi-mode DPN (again, definite photon-number) state, which generalizes the two-mode DPN state considered in Ref. \cite{u.doner}, and is written as
\begin{equation}\label{dpn}
    |\psi_N\rangle =\sum_{k_0+k_1+\cdots+k_m=N}\sqrt{\alpha_{\bm{k}}}|{\bm{k}}\rangle,
\end{equation}
where $|\bm{k}\rangle=|k_0\rangle\otimes|k_1\rangle\otimes\cdots\otimes|k_m\rangle$ denotes the multi-mode Fock states, $\alpha_{\bm{k}}$ are assumed to be non-negative, and the prior relative phases are all set to be zero without loss of generality. Note that here, the definite number $N$ of photons is distributed across $m$ modes. As described earlier, the DPN probe state of Eq.~(\ref{dpn}) undergoes the phase shift of $\hat{U}_{\bm {\widetilde{\phi}}}$ and subsequently the loss channels, resulting in the noisy phase-encoded DPN state prior to the measurement. It is written as $\hat{\rho}_{N,\bm{\phi}}=\left(\bigoplus_{l=0}^{N-1}\hat{\sigma}_{l,\bm{\phi}}\right)\oplus\hat{\sigma}_{N}$ where $\hat{\sigma}_{l,\bm{\phi}}$ is a (sub-normalized) statistical operator defined on a subspace spanned by $(N-l)$-photon multi-mode Fock states (see Supplementary Note I for the details). The QFIM of the noisy DPN states can thus be given as
\begin{equation}\label{QCRB_n_dpn}
    \mathbf{\bm{F}}_{\rm Q}\left(\hat{\rho}_{N,\bm{\phi}}\right)=\sum_{l=0}^{N-1}\mathbf{\bm{F}}_{\rm Q}\left(\hat{\sigma}_{l,\bm{\phi}}\right).
\end{equation}

Let us now compare the QCRB associated with the above QFIM to the SQL identified for the same estimation scenario. The SQL is written as \cite{m.namkung}
\begin{equation}\label{SQL_}
    \mathrm{Tr}\left\{\mathbf{\bm{F}}_{\rm Q,coh}^{-1}\right\}=\frac{1}{4N}\left\{\sqrt{\frac{m}{1-\gamma_0}}+\sum_{j=1}^{m}\frac{1}{\sqrt{1-\gamma_j}}\right\}^2,
\end{equation}
where $\mathbf{\bm{F}}_{\rm Q,coh}$ is a QFIM for a noisy phase-encoded multi-mode coherent state in the same setup shown in Fig.~\ref{fig1} \cite{m.namkung}. Note that the SQL written above is obtained for a single-mode coherent state input with a mean photon number $N$, which is fed into a multi-mode beam-splitter with optimized splitting ratios tailored to given loss rates, as demonstrated in Ref.~\cite{m.namkung}. Interestingly, the SQL equals the QCRB for a scenario where a single photon is fed into an optimized multi-mode beam-splitter, but with additional $N$ repetitions (see Methods and Supplementary Note II for the details regarding the repeated single photon scenario). In this case, $N$ single photons behave independently through the multi-mode beam-splitter and the photon loss channels, consequently enabling to interpret it as the case that employs $N$ distinguishable particles with the permutation of all ($N+m$)-dimensional labels for individual particles and modes (see Supplementary Note II for the details). It can finally be shown that, through the Bosonic symmetry inherent in second quantization \cite{a.peres}, the case of $N$ distinguishable particles is equivalent to the $N$-photon input scenario. Thus, the SQL is identical to the QCRB for a state generated by entering $N$-photon Fock state into single input port of a multi-mode beam-splitter. A simple example regarding the equivalences is presented in Methods.

Since the $N$-photon Fock state input represents an extreme case of the DPN state, it is evident that the QCRB for the DPN state encompasses the $N$-photon Fock state input. In other words, the QCRB for the DPN state input is always lower than or equal to the SQL of Eq.~(\ref{SQL_}), regardless of the loss rates. This conclusively addresses the primary question posed in the introduction. 

Moreover, compared to the case of NOON states in Ref.~\cite{m.namkung}, the QFIM of Eq.~(\ref{QCRB_n_dpn}) contains positive-definite QFIM terms with respect to $\hat{\sigma}_{l,\bm{\phi}}$ with $l>0$, which are absent in the NOON state input. It implies that the additional components in Eq.~(\ref{QCRB_n_dpn}) lead to additional non-vanishing QFIM terms and are useful for simultaneous multiple-phase estimation, thereby outperforming the NOON state schemes in lossy environment.

\begin{figure*}[t]
\centerline{\includegraphics[width=18cm]{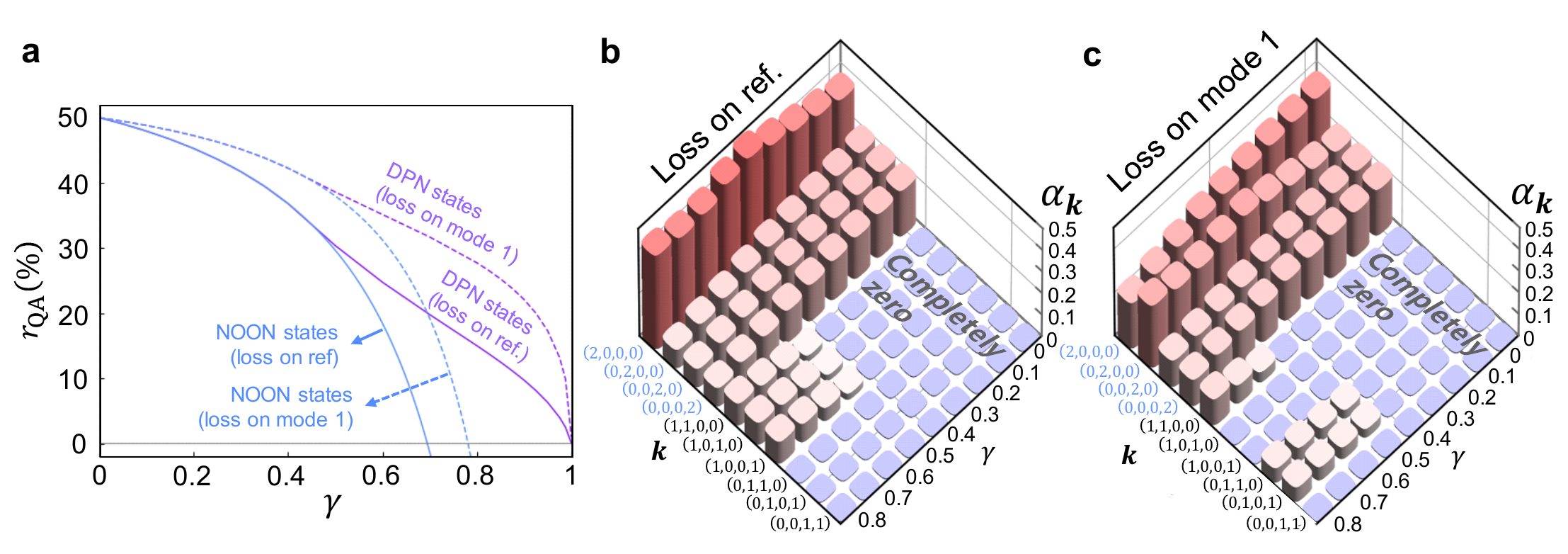}}
\caption{\textbf{Comparison of quantum advantage between the case of photon loss on a reference mode and that on mode 1.} \textbf{a} $r_{\rm QA}$ with respect to optimal four-mode two-photon DPN and NOON states, when either the reference mode or the mode 1 is subject to photon loss. The optimal weights of the DPN states shown in panel \textbf{a} are illustrated in \textbf{b} and \textbf{c}, respectively.}
\centering
\label{fig3}
\end{figure*}


Above, we have analytically demonstrated that the optimized DPN states consistently outperform the coherent state input, which serves as the classical benchmark, across all loss rates. However, the extent of the quantum advantage has not been provided due to difficulties in analytically calculating the inverse of the QFIM, as explained in Supplementary Note I. Here, we rather perform the numerical calculation of the QCRB of Eq.~(\ref{QCRB_n_dpn}) for the optimal DPN states to evaluate the quantum advantage over the SQL as well as the schemes employing the optimal NOON states from Ref.~\cite{m.namkung}. In this work, we exemplify simultaneous estimation of three phases using a probe state constrained to a mean photon number of $N=2$. In Fig.~\ref{fig2}\textbf{a}, we compare the QCRBs for the optimal DPN states, optimal NOON states, and the coherent state inputs when all photon loss rates are assumed to be equal to $\gamma$ \cite{c.oh,m.namkung}. It is clear that the scheme with the optimal DPN states outperforms the other two states, as expected from the proof sketch provided in Methods, further detail of which is also provided in Supplementary Note II. The quantum advantage over the SQL is maximized when $\gamma=0$, diminishes as $\gamma$ increases, and eventually disappears in the limit of $\gamma\rightarrow1$.  Additionally, it is observed that when $\gamma$ is zero or small, the optimal DPN state is almost identical to the optimal NOON state proposed in Refs.~\cite{p.c.humphreys,s.hong,s.hong2,m.namkung}. It means that the NOON states are still useful in environments with moderate photon loss ($\gamma\lesssim 0.34$). In Fig.~\ref{fig2}\textbf{b}, we further investigate the quantum advantages of the respective schemes using the DPN states and the NOON states compared to the SQL. Here, the quantum advantage is quantified by
\begin{equation}\label{QA}
    r_{\rm QA}=1-\frac{\mathrm{Tr}\left[\mathbf{F}_{\rm Q}^{-1}(\hat{\rho}_{N,\bm{\phi}})\right]}{\mathrm{Tr}\left[\mathbf{F}_{\rm Q,coh}^{-1}\right]},
\end{equation}
i.e., $r_{\rm QA}>0$ implies quantum advantage. Again, the quantum advantage for the DPN state input persists across all photon loss rates, whereas that for the NOON state input diminishes near $\gamma\approx 0.5$. Also, note that the maximal quantum advantage in the absence of loss is 50\% which is encapsulated by the theoretical value $(100/N)$\% in the lossless case. In Fig.~\ref{fig2}\textbf{c}, we illustrate the optimal population of the components comprising the DPN state with varying $\gamma$ in units of 0.1. It clearly shows that the optimal NOON states proposed in Refs. \cite{p.c.humphreys,m.namkung,j.urrehman,s.hong,s.hong2} are optimal when photon loss is moderate. As the photon loss rates increase, components other than the NOON states begin to contribute to the configuration of the optimal DPN state. The optimal weights exhibit the symmetry around the reference mode (i.e., the mode 0), such that $\alpha_{(0,2,0,0)}=\alpha_{(0,0,2,0)}=\alpha_{(0,0,0,2)}$, $\alpha_{(1,1,0,0)}=\alpha_{(1,0,1,0)}=\alpha_{(1,0,0,1)}$, and  $\alpha_{(0,1,1,0)}=\alpha_{(0,1,0,1)}=\alpha_{(0,0,1,1)}$. This is because all photon loss rates are assumed to be equal, and the common reference mode is set in the estimation scenarios. It is also evident that the optimal weights are more concentrated on the reference mode since the reference phase is used $m$ times to estimate all the relative phases.  

The aforementioned asymmetry between the reference mode and the other modes in our estimation configuration implies that photon loss on the reference may deteriorate the estimation capability more significantly compared to loss on the other modes. This issue is absent in single-phase estimation problems, where the reference and the signal modes are symmetric to each other \cite{l.hwang,u.doner}. Thus, it is particularly important to compare the effect of loss on the respective modes, i.e., either the reference or the others in our scenario,  to understand the unbalanced feature of multiple-phase estimation. Therefore, we compare two cases: one where the loss with $\gamma$ occurs on the reference mode and another where the loss with $\gamma$ occurs on the mode 1. Figure \ref{fig3}\textbf{a} shows their difference in terms of the quantum advantage ratio $r_{\rm QA}$ of Eq.~(\ref{QA}). This clearly demonstrates that when the common reference mode is subject to loss, the quantum advantage is lower than the other case. In Figs.~\ref{fig3}\textbf{b} and \textbf{c}, we also elaborate on the optimal weights of the DPN states considered in Fig.~\ref{fig3}\textbf{a}. When the reference mode is only exposed to the lossy environment, the similar weights distribution appears as in Fig.~\ref{fig2}\textbf{c}. Meanwhile when the mode 1 is only lossy, more energy is needed to compensate for the loss on the mode 1.

\begin{figure*}[t]
\centerline{\includegraphics[width=18cm]{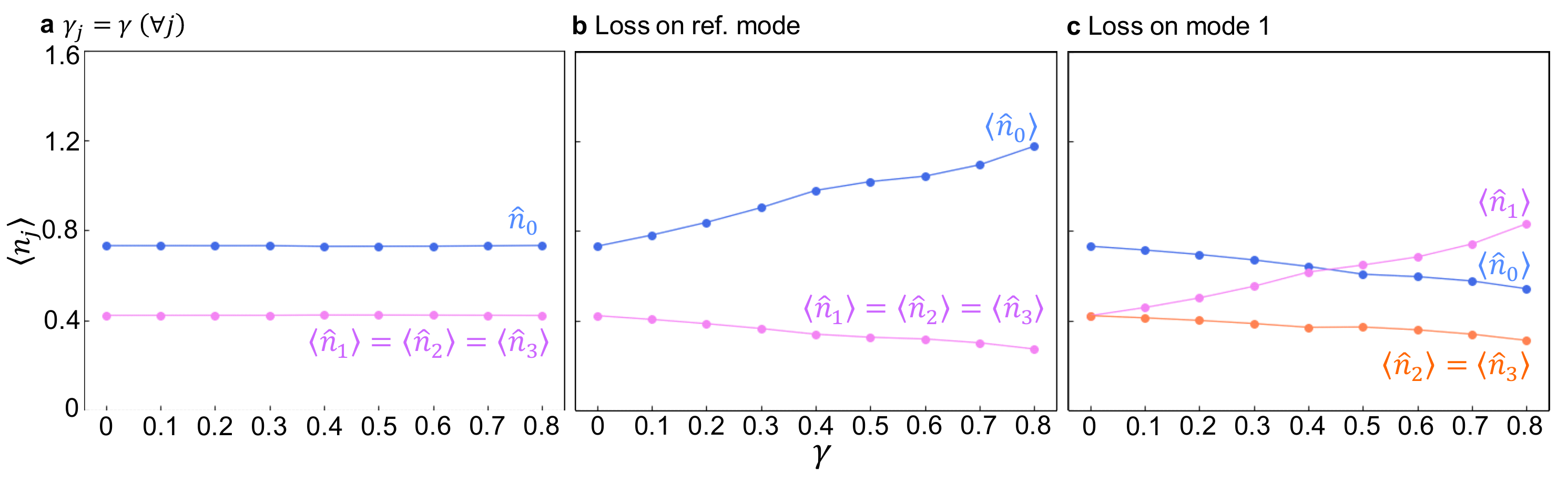}}
\caption{\textbf{Energies $\bm{\langle \hat{n}_0\rangle}$, $\bm{\langle \hat{n}_1\rangle}$, $\bm{\langle \hat{n}_2\rangle}$, and $\bm{\langle \hat{n}_3\rangle}$ with respect to photon loss rate $\bm{\gamma\in[0,0.8]}$.} \textbf{a} In case that all photon loss rates are equal to $\gamma$. \textbf{b} In case that only a reference mode is affected by photon loss with loss rate $\gamma$. \textbf{c} In case that the mode 1 is affected by the photon loss.}
\centering
\label{fig4}
\end{figure*}

For the results shown in Figs.~\ref{fig2}\textbf{c},~\ref{fig3}\textbf{b}, and~\ref{fig3}\textbf{c}, we also evaluate the average photon numbers of the individual modes across different cases. When an identical amount of photon loss occurs to all the modes, the average photon number distribution over the modes does not vary with the loss rate as shown in Fig.~\ref{fig4}\textbf{a}. There, the reference mode uses more energy than the other modes since it is used $m$ times in the simultaneous estimation. When only the reference mode is exposed to the photon loss, significantly more energy is needed on the reference mode and an increasing behaviour  with the loss rate is observed, as shown in Fig.~\ref{fig4}\textbf{b}. This behaviour is reversed when photon loss occurs only on the mode 1. Figure~\ref{fig4}\textbf{c} shows that the reference mode uses the largest amount of energy when the loss is small, but the mode 1 starts to require more energy across $\gamma \approx 0.45$. Again, more energy is generally required to reduce the uncertainty on the mode where photon loss occurs.  We also note that the tendency of $\langle \hat{n}_0\rangle$ well matches from $\alpha_{(2,0,0,0)}$ in Fig.~\ref{fig4}\textbf{b}, and similarly, $\langle \hat{n}_1\rangle$ reflects $\alpha_{(0,2,0,0)}$ in Fig.~\ref{fig4}\textbf{c}.

\section*{Discussion}
We have demonstrated that an optimal multiple-phase estimation scheme using multi-mode DPN states provides persistent quantum advantage over the SQL, regardless of photon loss levels. When photon loss is moderate, the optimal DPN states reduce to the optimal NOON states. This means that the NOON states proposed in Ref.~\cite{m.namkung} are still useful when photon loss is mild. However, as the loss increases, additional components absent in the NOON state begin to contribute to enhancing the estimation precision, forming the optimal DPN states. We also show that the reference mode is the most sensitive to photon loss compared to the other modes, as it is used $m$ times for estimating $m$ relative phases. Furthermore, we have examined the impact of photon loss in individual modes, demonstrating that additional energy is required where loss occurs. This suggests that average photon numbers should be rebalanced according to specific loss rates when forming optimal DPN states.

We believe that our results contribute to the development of quantum sensing technology, demonstrating that quantum enhancement therein is achievable even in noisy environments. We emphasize that our proposed methodology is broadly applicable to various multiple-phase estimation scenarios, promising for future exploration. Unlike schemes relying on error mitigation or post-selection, our scheme effectively suppresses the effect of photon loss while efficiently utilizing available resources. It is also interesting to employ the DPN states not only for simultaneous sensing of multiple parameters such as two-photon absorption parameters \cite{a.karsa}, distributed phases \cite{x.guo,l.-z.liu,d.-h.kim}, and incompatible parameters \cite{f.hanamura,f.hanamura2}, but also in estimating evolution time of a non-Hermitian \cite{x.yu} and open quantum systems \cite{a.das}. We have considered a specific reference configuration where one of many modes is focused to serve as a reference. Thus, it is also a meaningful direction to consider other reference configurations \cite{a.z.goldberg}.

\section*{Methods}

\begin{figure}[t]
\centerline{\includegraphics[width=8.5cm]{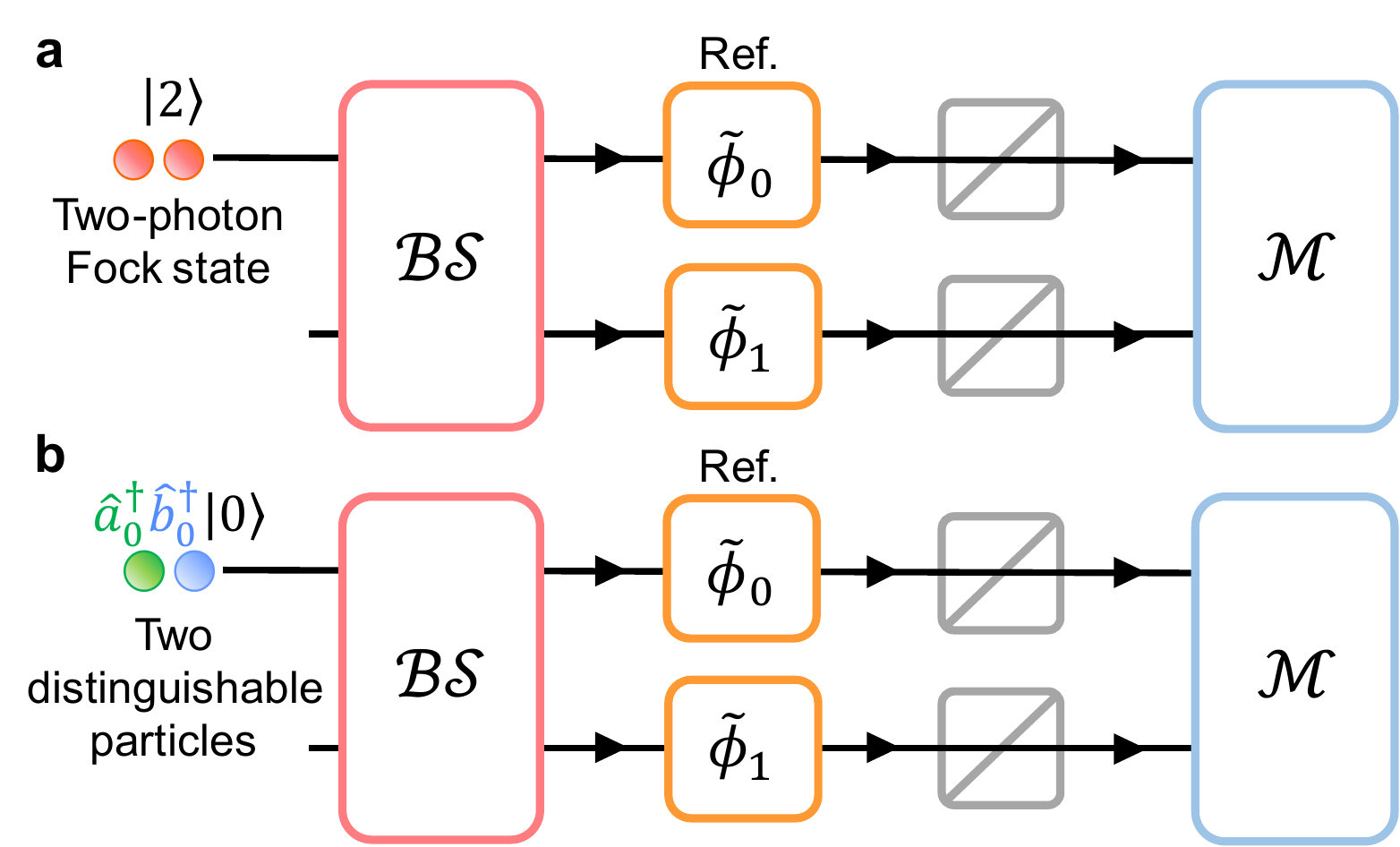}}
\caption{\textbf{Single-phase estimation scheme for estimating an unknown relative phase $\bm{\phi=\widetilde{\phi}_1-\widetilde{\phi}_0}$.} \textbf{a} In case that a two-photon Fock state $|2\rangle$ is exploited. \textbf{b} In case that two distinguishable photons are exploited. In these figures, $\mathcal{BS}$ and $\mathcal{M}$ denote a beam-splitter and a measurement, respectively.}
\centering
\label{fig5}
\end{figure}

{\bf Example of Bosonic symmetry for equivalence between SQL and QCRB of $N$-photon Fock state input.} We first note that each single photon undergoing a multi-mode beam-splitter is equal to an one-photon NOON state. This fact is used for showing that the SQL is equal to the QCRB of single photon input with $N$ repetitions. Here, we provide an example demonstrating the equivalence between a two-photon Fock state and two distinguishable particles in estimating a single relative phase as described in Fig.~\ref{fig5}. \textit{The detailed version of this Method with the arbitrary number of photons $N$ and modes $m$ is provided by Supplementary Note II.}

Let us first consider in Fig.~\ref{fig5}\textbf{a} that a two-photon Fock state is distributed by a beam-splitter in probe state generation. Here the beam-splitter transforms the two-photon Fock state input to
\begin{equation}
    |2\rangle\rightarrow \sqrt{\alpha_0}|20\rangle+\sqrt{\alpha_1}|11\rangle+\sqrt{\alpha_2}|02\rangle,
\end{equation}
with $\alpha_0=u_0^2$, $\alpha_1=2u_0u_1$, and $\alpha_2=u_1^2$, 
where $u_0$ and $u_1$ are assumed to be non-negative real numbers such that $u_0^2+u_1^2=1$ without loss of generality. After undergoing phase shifts, the phase-encoded state is obtained as
\begin{equation}\label{Focks}
    |\psi_{\bm{\phi}}^{\rm (Fock)}\rangle = \sqrt{\alpha_0}|20\rangle+\sqrt{\alpha_1}e^{i\phi}|11\rangle+\sqrt{\alpha_2}e^{2i\phi}|02\rangle.
\end{equation}
Let a measurement employed for extracting phase information be represented as a POVM $\mathcal{M}=\left\{\hat{M}_l:l\in\Omega\right\}$ with respect to outcomes $l$ in an outcome space $\Omega$. From the Stinespring dilation theorem, the measurement probability is described by
\begin{equation}\label{prob_mes}
    p(l|\bm{\phi},\mathrm{Fock})=\langle\psi_{\bm{\phi}}^{\rm (Fock)}|\hat{\Pi}_{l}|\psi_{\bm{\phi}}^{\rm (Fock)}\rangle,
\end{equation}
where $\hat{\Pi}_l$ is an operator defined on signal mode system as 
\begin{equation}
    \hat{\Pi}_l={}_{\rm E}\langle0|\hat{U}_{\bm{\gamma}}^{\dagger}\left(\hat{M}_l\otimes\hat{\mathbb{I}}_{\rm E}\right)\hat{U}_{\bm{\gamma}}|0\rangle_{\rm E}
\end{equation}
with an identity operator $\hat{\mathbb{I}}_{\rm E}$ on an environment system, a fictitious vacuum state $|0\rangle_{\rm E}$ on the environment, and a unitary operator $\hat{U}_{\bm{\gamma}}$ that describes the photon loss \cite{d.mcmahon}. It is straightforward to decompose $\hat{\Pi}_l$ as
\begin{equation}
    \hat{\Pi}_l=\sum_{a,b\in\{20,11,02\}}\Pi_{l,ab}|a\rangle\langle b|+\sum_{a,b\not\in\{20,11,02\}}\Pi_{l,ab}|a\rangle\langle b| 
\end{equation}
with complex numbers $\Pi_{l,ab}$ in general, and the probability in Eq.~(\ref{prob_mes}) as
\begin{equation}
    p(l|\bm{\phi},\mathrm{Fock})=\sum_{a,b\in\{20,11,02\}}\alpha_a\Pi_{l,ab}\alpha_b.
\end{equation}

In Fig.~\ref{fig5}\textbf{b}, it is considered that two distinguishable photons, which are represented by two mode operators $\hat{a}_0$ and $\hat{b}_0$, enter the input mode 0 of the beam-splitter. Then, the state of these two distinguishable photons distributed by the beam-splitter and undergoing the phase shifts is described as
\begin{equation}
    |\psi_{\bm{\phi}}^{\rm (dist.)}\rangle = \sqrt{\alpha_0}|\widetilde{20}\rangle+\sqrt{\alpha_1}e^{i\phi}|\widetilde{11}\rangle+\sqrt{\alpha_2}e^{2i\phi}|\widetilde{02}\rangle,
\end{equation}
where $\alpha_j$ are equal to what we introduced in Eq.~(\ref{Focks}). Here, $|\widetilde{20}\rangle$, $|\widetilde{11}\rangle$, and $|\widetilde{02}\rangle$ are defined as 
\begin{eqnarray}
    |\widetilde{20}\rangle&=&\hat{a}_0^\dagger\hat{b}_0^\dagger|0\rangle,\nonumber\\
    |\widetilde{11}\rangle&=&\frac{1}{\sqrt{2}}\left(\hat{a}_0^\dagger\hat{b}_1^\dagger+\hat{a}_1^\dagger\hat{b}_0^\dagger\right)|0\rangle,\nonumber\\
    |\widetilde{02}\rangle&=&\hat{a}_1^\dagger\hat{b}_1^\dagger|0\rangle,
\end{eqnarray}
with mode operators $\hat{a}_j$ and $\hat{b}_j$ which are defined on the mode $j$. We note that these vectors have Bosonic symmetry \cite{a.peres}. For given operators $\hat{\Pi}_l$ induced by the POVM $\mathcal{M}$, the measurement probability is derived as
\begin{equation}
    p(l|\bm{\phi},\mathrm{dist.})=\langle\psi_{\bm{\phi}}^{\rm (dist.)}|\hat{\Pi}_{l}|\psi_{\bm{\phi}}^{\rm (dist.)}\rangle.
\end{equation}
Let us consider an isometry $\hat{V}$
\begin{eqnarray}
    \hat{V}=|\widetilde{20}\rangle\langle 20|+|\widetilde{11}\rangle\langle 11|+|\widetilde{02}\rangle\langle 02|,
\end{eqnarray}
which performs transformation ${span}\left\{|20\rangle,|11\rangle,|02\rangle\right\}\rightarrow{span}\left\{|\widetilde{20}\rangle,|\widetilde{11}\rangle,|\widetilde{02}\rangle\right\}$. This isometry can also satisfy $\hat{V}|\psi_{\bm{\phi}}^{\rm (Fock)}\rangle=|\psi_{\bm{\phi}}^{\rm (dist.)}\rangle$, and thus
\begin{eqnarray}\label{eq_res}
    p(l|\bm{\phi},\mathrm{dist.})&=&\langle\psi_{\bm{\phi}}^{\rm (Fock)}|\hat{V}\hat{\Pi}_l\hat{V}^{\dagger}|\psi_{\bm{\phi}}^{\rm (Fock)}\rangle\nonumber\\
    &=&\sum_{a,b\in\{20,11,02\}}\alpha_a\Pi_{l,ab}\alpha_b\nonumber\\
    &=&p(l|\bm{\phi},\mathrm{Fock})
\end{eqnarray}
which shows the equivalence between the two-photon Fock state and the two distinguishable photons in a single-phase estimation. By the virtue of the equivalence relation in Eq.~(\ref{eq_res}), we eventually conclude that both states provide the same minimum QCRB over all possible beam-splitters, which leads us to the conclusion that the SQL is equal to the QCRB of the Fock states (The detailed proof regarding the most general case is provided in Supplementary Note II).

\begin{widetext}
    \section*{Supplementary Note I: Evaluation of quantum Fisher information matrix (QFIM)}
Let us begin the explanation by recalling phase-encoded multi-mode definite photon-number (DPN) states
\begin{align}
    |\psi_{N,\bm{\phi}}\rangle&=\sum_{k_0+\cdots+k_m=N}\alpha_{k_0\cdots k_m}e^{i(k_1\phi_1+\cdots+k_m\phi_m)}|k_0\rangle_{\hat{a}_0}\otimes|k_1\rangle_{\hat{a}_1}\otimes\cdots\otimes|k_m\rangle_{\hat{a}_m}\nonumber\\
    &=\sum_{k_0+\cdots+k_m=N}\frac{\alpha_{k_0\cdots k_m}e^{i(k_1\phi_1+\cdots+k_m\phi_m)}}{\sqrt{k_0!k_1!\cdots k_m!}}\hat{a}_0^{\dagger k_0}\hat{a}_1^{\dagger k_1}\cdots\hat{a}_m^{\dagger k_m}|0\rangle_{\hat{a}_0\cdots\hat{a}_m},
\end{align}
where $\hat{a}_j$ is a mode operator defined on the mode $j$. The mode transformations $\hat{a}_j\rightarrow\sqrt{1-\gamma_j}\hat{a}_j+\sqrt{\gamma_j}\hat{e}_j$ together with environment mode operators $\hat{e}_j$, which represent photon loss channels, leads us to a composite state between the phase-encoded probe and environment
\begin{align}
    |\Psi_{N,\bm{\gamma},\bm{\phi}}\rangle&=\sum_{k_0+\cdots+k_m=N}\frac{\alpha_{k_0\cdots k_m}e^{i(k_1\phi_1+\cdots+k_m\phi_m)}}{\sqrt{k_0!k_1!\cdots k_m!}}\left\{\prod_{j=0}^{m}(\sqrt{1-\gamma_j}\hat{a}_j^{\dagger}+\sqrt{\gamma_j}\hat{e}_j^{\dagger})^{k_j}\right\}|0\rangle_{\hat{a}_0\cdots\hat{a}_d}\otimes|0\rangle_{\hat{e}_0\cdots\hat{e}_m}\nonumber\\
    &=\sum_{k_0+\cdots+k_m=N}\alpha_{k_0\cdots k_m}e^{i(k_1\phi_1+\cdots+k_m\phi_m)}\sum_{l_j\le k_j}\left\{\prod_{j=0}^{m}\sqrt{\frac{k_j!}{l_j!(k_j-l_j)!}}(1-\gamma_j)^{\frac{k_j-l_j}{2}}\gamma_j^{\frac{l_j}{2}}\right\}\left\{\bigotimes_{s=0}^{m}|k_{s}-l_{s}\rangle_{\hat{a}_s}\otimes|l_{s}\rangle_{\hat{e}_s}\right\}.
\end{align}
The lossy phase-encoded DPN state is evaluated by tracing the above state over all environment modes as
\begin{align}\label{den_detail}
    \hat{\rho}_{N,\bm{\phi}}&=\mathrm{Tr}_{\hat{e}_0\cdots\hat{e}_m}\left\{|\Psi_{N,\bm{\gamma},\bm{\phi}}\rangle\langle\Psi_{N,\bm{\gamma},\bm{\phi}}|\right\}\nonumber\\
    &=\sum_{\substack{k_0+\cdots+k_m=N \ \wedge \\  k_0'+\cdots+k_m'=N}}\alpha_{\bm{k}}\alpha_{\bm{k}'}e^{i(k_1-k_1')\phi_1+\cdots+i(k_m-k_m')\phi_m}\sum_{0\le l_j\le \min\{k_j,k_{j}'\} \ \forall j}P_{\bm{k},\bm{l}}P_{\bm{k}',\bm{l}}\left\{\bigotimes_{s=0}^{m}|k_s-l_s\rangle \langle k_{s}'-l_s|\right\}\nonumber\\
    &=\sum_{\substack{0\le l_j\le N \ \forall j \\ l_0+\cdots+l_m\le N}}\sum_{\substack{k_j\ge l_j \ \forall j \\ k_0+\cdots+k_m=N}}\sum_{\substack{k_j'\ge l_j \ \forall j \\ k_0'+\cdots+k_m'=N}}\alpha_{\bm{k}}\alpha_{\bm{k}'}e^{i(k_1-k_1')\phi_1+\cdots+i(k_m-k_m')\phi_m}P_{\bm{k},\bm{l}}P_{\bm{k}',\bm{l}}\left\{\bigotimes_{s=0}^{m}|k_s-l_s\rangle \langle k_{s}'-l_s|\right\},
\end{align}
where $P_{\bm{k},\bm{l}}$ is defined as
\begin{equation}
    P_{\bm{k},\bm{l}}=\sqrt{\frac{k_0!\cdots k_m!}{l_0!(k_0-l_0)!\cdots l_m!(k_m-l_m)!}}(1-\gamma_0)^{\frac{k_0-l_0}{2}}\gamma_0^{\frac{l_0}{2}}\cdots(1-\gamma_d)^{\frac{k_m-l_m}{2}}\gamma_m^{\frac{l_m}{2}}.
\end{equation}
Let us write a non-negative parameter $w_{\bm{l}}$ and a corresponding phase-dependent vector $|w_{\bm{l},\bm{\phi}}\rangle$ as
\begin{align}
    w_{\bm{l}}&=\sum_{\substack{k_j\ge l_j \ \forall j \\ k_0+\cdots+k_m=N}}\alpha_{\bm{k}}^2P_{\bm{k},\bm{l}}^2, \nonumber\\ 
    |w_{\bm{k},\bm{l}}\rangle&=\frac{1}{\sqrt{w_{\bm{l}}}}\sum_{\substack{k_j\ge l_j \ \forall j \\ k_0+\cdots+k_m=N}}\alpha_{\bm{k}}P_{\bm{k},\bm{l}}e^{i(k_1\phi_1+\cdots+k_m\phi_m)}|k_0-l_0\rangle_{\hat{a}_0}\otimes\cdots\otimes|k_m-l_m\rangle_{\hat{a}_m}.
\end{align}
Then, the density operator written in Eq.~(\ref{den_detail}) is concisely decomposed as
\begin{equation}\label{decom}
    \hat{\rho}_{N,\bm{\phi}}=\sum_{\substack{0\le l_j\le N \ \forall j \\ l_0+\cdots+l_m\le N}}w_{\bm{l}}|w_{\bm{l},\bm{\phi}}\rangle\langle w_{\bm{l},\bm{\phi}}|=\sum_{l=0}^{N}\sum_{\substack{0\le l_j\le N \ \forall j \\ l_0+\cdots+l_m= N-l}}w_{\bm{l}}|w_{\bm{l},\bm{\phi}}\rangle\langle w_{\bm{l},\bm{\phi}}|=\sum_{l=0}^{N}\hat{\sigma}_{l,\bm{\phi}}.
\end{equation}
We note that any two subspaces $\mathrm{span}\{|w_{\bm{l},\bm{\phi}}\rangle:l_0+\cdots+l_m=N-l\}$ and $\mathrm{span}\{|w_{\bm{l},\bm{\phi}}\rangle:l_0+\cdots+l_m=N-l'\}$ are orthogonal to each other. Thus, two supports $\mathrm{supp}(\hat{\sigma}_{\bm{l},\bm{\phi}})$ and $\mathrm{supp}(\hat{\sigma}_{\bm{l'},\bm{\phi}})$ are also orthogonal, and thus the decomposition in Eq.~(\ref{decom}) is equivalently described as $\hat{\rho}_{N,\bm{\phi}}=\bigoplus_{l=0}^{N}\hat{\sigma}_{l,\bm{\phi}}$. Finally, the orthgonality between these subspaces leads us to 
\begin{equation}\label{qfim_atlast}
    \mathbf{F}_{\rm Q}\left(\hat{\rho}_{N,\bm{\phi}}\right)=\sum_{l=0}^{N}\mathbf{F}_{\rm Q}\left(\hat{\sigma}_{l,\bm{\phi}}\right).
\end{equation}
If $l=0$, then the corresponding statistical operator is a sub-normalized pure state $\hat{\sigma}_{0,\bm{\phi}}=|v_{0,\bm{\phi}}\rangle\langle v_{0,\bm{\phi}}|$ and corresponding QFIM elements is evaluated as
\begin{align}\label{qfim_sub1}
    \mathbf{F}_{\mathrm{Q}(j,k)}\left(\hat{\sigma}_{0,\bm{\phi}}\right)=4\mathrm{Re}\left\{\langle\partial_jv_{0,\bm{\phi}}|\partial_kv_{0,\bm{\phi}}\rangle-\langle\partial_jv_{0,\bm{\phi}}|v_{0,\bm{\phi}}\rangle\langle v_{0,\bm{\phi}}|\partial_kv_{0,\bm{\phi}}\rangle\right\}.
\end{align}
Otherwise, the QFIM of $\hat{\sigma}_{l,\bm{\phi}}$ has rank more than or equal to two, and its QFIM element is 
\begin{align}\label{qfim_sub2}
    \mathbf{F}_{{\rm Q}(j,k)}\left(\hat{\sigma}_{l,\bm{\phi}}\right)=2\mathrm{vec}(\partial_j\hat{\sigma}_{l,\bm{\phi}})^\dagger \left(\hat{\sigma}_{l,\bm{\phi}}\otimes\hat{\mathbb{I}}+\hat{\mathbb{I}}\otimes\hat{\sigma}_{l,\bm{\phi}}^*\right)^{-1}\mathrm{vec}(\partial_k\hat{\sigma}_{l,\bm{\phi}}),
\end{align}
where $\hat{\mathbb{I}}$ is an identity operator and $\mathrm{vec}(\partial_j\hat{\sigma}_{l,\bm{\phi}})$ is a vector in Liouville space with respect to $\partial_j\hat{\sigma}_{l,\bm{\phi}}$ \cite{d.safranek}. The inverse of the QFIM in Eq. (\ref{qfim_atlast}) is not analytically derived in general due to the difficulty in calculating the inverse of $\hat{\sigma}_{l,\bm{\phi}}\otimes\hat{\mathbb{I}}+\hat{\mathbb{I}}\otimes\hat{\sigma}_{l,\bm{\phi}}^*$ in Eq.~(\ref{qfim_sub2}). Thus, we need to perform the numerical method to evaluate the minimum QCRB over all DPN states. If $\hat{\rho}_{N,\bm{\phi}}$ becomes singular while iteratively minimizing the QCRB, then the singularity may spoil the evaluation of the inverse in Eq. (\ref{qfim_sub2}). There is one way to tame the singularity by redefining the QCRB in terms of the Moore-Penrose pseudoinverse \cite{a.ben,j.xavier}, but it occasionally provides overestimated precision bound \cite{a.z.goldberg0}. Thus, in order to avoid such the unexpected obstacle, we alternatively let the statistical operators be $\hat{\sigma}_{l,\bm{\phi}}\rightarrow(1-\nu)\hat{\sigma}_{l,\bm{\phi}}+\nu\frac{\hat{\mathbb{I}}}{\mathrm{dim}(\hat{\sigma}_{l,\bm{\phi}})}$ with sufficiently small $\nu>0$, as what is recommended by Ref. \cite{d.safranek}. 

\section*{Supplementary Note II: Equivalence between SQL and QCRB of $N$-photon Fock state input (Generalized version of Methods)}
For the direct approach of the proof, we need to evaluate all possible cases of an $N$-photon Fock state entering a multi-mode beam-splitter. However, this approach is related to a Boson sampling problem, which is known to be difficult to solve \cite{s.aaronson,c.s.hamilton}. Thus, for the simpler approach, we alternatively show that the case of the $N$-photon Fock state is transformed to the case of $N$ distinguishable particles in multiple-phase estimation problem.

 We first remind that an $N$-photon Fock state $|N\rangle$ after passing a multi-mode beam-splitter $\hat{U}_{\rm BS}$ and phase-shifters $\hat{U}_{\bm{\phi}}$ is described as
\begin{equation}
    |\psi_{N,\bm{\phi}}^{\rm (Fock)}\rangle=\hat{U}_{\bm{\phi}}\hat{U}_{\rm BS}|N\rangle_{\hat{a}_0}=\sum_{k_0+\cdots+k_m=N}\sqrt{\frac{N!}{k_0!\cdots k_m!}}u_0^{k_0}\cdots u_m^{k_m}e^{i(k_1\phi_1+\cdots+k_m\phi_m)}|k_0\rangle_{\hat{a}_0}\otimes|k_1\rangle_{\hat{a}_1}\cdots\otimes|k_m\rangle_{\hat{a}_m},
\end{equation}
with positive real numbers $u_0,\cdots,u_m$ such that $u_0^2+\cdots+u_m^2=1$ without loss of generality. Let us also introduce $N$ distinguishable photons \cite{m.gessner}
\begin{equation}
   |\psi_{N}^{\rm (dist.)}\rangle=\prod_{j=1}^{N}\hat{a}_0^{(r)\dagger}|0\rangle_{\hat{a}_0},
\end{equation}
where $\hat{a}_{0}^{(r)}$ is a mode 0 operator for representing $r$-th particle. The phase-encoded $N$ distinguishable particles are fully described as
\begin{align}
     |\psi_{N,\bm{\phi}}^{\rm (dist.)}\rangle=\hat{U}_{\bm{\phi}}\hat{U}_{\rm BS}|\psi_{N}^{\rm (dist.)}\rangle_{\hat{a}_0}=\sum_{k_0+\cdots+k_m=N}\sqrt{\frac{N!}{k_0!\cdots k_m!}}u_0^{k_0}\cdots u_m^{k_m}e^{i(k_1\phi_1+\cdots+k_m\phi_m)}|v_{(k_0,\cdots,k_m)}\rangle_{\hat{a}_0\cdots\hat{a}_m},
\end{align}
where $|v_{(k_0,\cdots,k_m)}\rangle$ is
\begin{align}
    |v_{(k_0,\cdots,k_m)}\rangle=\frac{1}{\sqrt{\frac{N!}{k_0!\cdots k_m!}}}\sum_{(r_1,\cdots,r_N)\in\mathcal{C}_N}\left(\hat{a}_0^{(r_1)\dagger}\cdots\hat{a}_{0}^{(r_{k_0})\dagger}\right)\left(\hat{a}_1^{(r_{k_0+1})\dagger}\cdots\hat{a}_{1}^{(r_{k_0+k_1})\dagger}\right)\cdots\left(\hat{a}_m^{(N-r_{k_m}+1)\dagger}\cdots\hat{a}_{m}^{(r_{N})\dagger}\right)|0\rangle
\end{align}
defined with a set $\mathcal{C}_N$ of all possible combinations of natural numbers $\{1,2,\cdots,N\}$. It is straightforward to show that $|k_0\rangle\otimes|k_1\rangle\otimes\cdots|k_m\rangle$ is one-to-one correspond to $|v_{(k_0,\cdots,k_m)}\rangle$. Thus, there exists an isometry $\hat{V}$ such that $\hat{V}|k_0\rangle\otimes\cdots\otimes|k_m\rangle=|v_{(k_0,\cdots,k_m)}\rangle$ and $\hat{V}^\dagger|v_{(k_0,\cdots,k_m)}\rangle=|k_0\rangle\otimes\cdots\otimes|k_m\rangle$. It also means that the phase-encoded $N$-distinguishable particles are transformed from the phase-encoded $N$-photon Fock state as $|\psi_{N,\bm{\phi}}^{\rm (dist.)}\rangle=\hat{V}|\psi_{N,\bm{\phi}}^{\rm (Fock)}\rangle$.

We show that the transformation $\hat{V}$ does not change the Fisher information matrix (FIM). Let us consider a POVM $\mathcal{M}=\left\{\hat{M}_l:l\in\Omega\right\}$ with outcomes $l$ in an outcome space $\Omega$, which is used for extracting phase information. Then, the FIM of the phase-encoded $N$-photon Fock state $|\psi_{N,\bm{\phi}}^{\rm (Fock)}\rangle$ under the photon loss is dependent on the measurement probability 
\begin{align}
    p(l|\phi,\mathrm{Fock})&=\mathrm{Tr}_{\hat{a}_0\cdots\hat{a}_m}\left\{\bigotimes_{j=0}^{m}\Lambda_{\gamma_j}\left(|\psi_{N,\bm{\phi}}^{\rm (Fock)}\rangle\langle\psi_{N,\bm{\phi}}^{\rm (Fock)}|\right)\hat{M}_l\right\}\nonumber\\
    &=\mathrm{Tr}_{\hat{a}_0\cdots\hat{a}_m}\left\{\mathrm{Tr}_{\hat{e}_0\cdots\hat{e}_m}\left(\hat{U}_{\bm{\gamma}}|\psi_{N,\bm{\phi}}^{\rm (Fock)}\rangle\langle\psi_{N,\bm{\phi}}^{\rm (Fock)}|_{\hat{a}_0\cdots\hat{a}_m}\otimes|0\rangle\langle 0|_{\hat{e}_0\cdots\hat{e}_m}\hat{U}_{\bm{\gamma}}^\dagger\right)\hat{M}_l\right\}\nonumber\\
    &=\mathrm{Tr}_{\hat{a}_0\cdots\hat{a}_m\hat{e}_0\cdots\hat{e}_m}\left\{\left(\hat{U}_{\bm{\gamma}}|\psi_{N,\bm{\phi}}^{\rm (Fock)}\rangle\langle\psi_{N,\bm{\phi}}^{\rm (Fock)}|_{\hat{a}_0\cdots\hat{a}_m}\otimes|0\rangle\langle 0|_{\hat{e}_0\cdots\hat{e}_m}\hat{U}_{\bm{\gamma}}^\dagger\right)\hat{M}_l\otimes\hat{\mathbb{I}}_{\hat{e}_0\cdots\hat{e}_m}\right\}\nonumber\\
    &=\langle\psi_{N,\bm{\phi}}^{\rm (Fock)}|_{\hat{a}_0\cdots\hat{a}_m}\otimes\langle0|_{\hat{e}_0\cdots\hat{e}_m}\hat{U}_{\bm{\gamma}}^\dagger \left(\hat{M}_l\otimes\hat{\mathbb{I}}_{\hat{e}_0\cdots\hat{e}_m}\right)\hat{U}_{\bm{\gamma}}|\psi_{N,\bm{\phi}}^{\rm (Fock)}\rangle_{\hat{a}_0\cdots\hat{a}_m}\otimes|0\rangle_{\hat{e}_0\cdots\hat{e}_m}\nonumber\\
    &=\langle\psi_{N,\bm{\phi}}^{\rm (Fock)}|\hat{\Pi}_{\bm{\gamma},\bm{l}}|\psi_{N,\bm{\phi}}^{\rm (Fock)}\rangle
\end{align}
with an operator defined on the signal modes $\hat{a}_0,\cdots,\hat{a}_m$:
\begin{eqnarray}
    \hat{\Pi}_{\bm{\gamma},\bm{l}}=\left(\hat{\mathbb{I}}_{\hat{a}_0\cdots\hat{a}_m}\otimes\langle0|_{\hat{e}_0\cdots\hat{e}_m}\right)\hat{U}_{\bm{\gamma}}^\dagger \left(\hat{M}_l\otimes\hat{\mathbb{I}}_{\hat{e}_0\cdots\hat{e}_m}\right)\hat{U}_{\bm{\gamma}}\left(\hat{\mathbb{I}}_{\hat{a}_0\cdots\hat{a}_m}\otimes|0\rangle_{\hat{e}_0\cdots\hat{e}_m}\right),
\end{eqnarray}
where $\hat{U}_{\bm{\gamma}}$ is an unitary operator on entire Hilbert space for representing photon loss with loss rates $\bm{\gamma}=(\gamma_0,\cdots,\gamma_m)$ \cite{d.mcmahon}. By using the isometry $\hat{V}$ discussed above, the measurement probability is rewritten as $p(l|\phi,\mathrm{Fock})=\langle\psi_{N,\bm{\phi}}^{\rm (dist.)}|\hat{V}\hat{\Pi}_{\bm{\gamma},\bm{l}}\hat{V}^\dagger|\psi_{N,\bm{\phi}}^{\rm (dist.)}\rangle$. We note that $\hat{V}$ just changes the computational basis consisting $\hat{\Pi}_{\bm{\gamma},\bm{l}}$, and does not change the components therein. It means that, for a given POVM $\mathcal{M}$, the measurement probability induced from the $N$-photon Fock state is equal to that from $N$ distinguisble particles $p(k|\phi,\mathrm{dist.})$, whereby two FIMs $\mathbf{F}\left(\bigotimes_{j=0}^{m}\Lambda_{\gamma_j}\left(|\psi_{N,\bm{\phi}}^{\rm (Fock)}\rangle\langle\psi_{N,\bm{\phi}}^{\rm (Fock)}|\right),\mathcal{M}\right)$ and $\mathbf{F}\left(\bigotimes_{j=0}^{m}\Lambda_{\gamma_j}\left(|\psi_{N,\bm{\phi}}^{\rm (dist.)}\rangle\langle\psi_{N,\bm{\phi}}^{\rm (dist.)}|\right),\mathcal{M}\right)$ are also equal to each other. Since there is a POVM such that FIM attains the QFIM, we obtain
\begin{equation}
    \mathbf{F}_{\rm Q}\left(\bigotimes_{j=0}^{m}\Lambda_{\gamma_j}\left(|\psi_{N,\bm{\phi}}^{\rm (Fock)}\rangle\langle\psi_{N,\bm{\phi}}^{\rm (Fock)}|\right)\right)=\mathbf{F}_{\rm Q}\left(\bigotimes_{j=0}^{m}\Lambda_{\gamma_j}\left(|\psi_{N,\bm{\phi}}^{\rm (dist.)}\rangle\langle\psi_{N,\bm{\phi}}^{\rm (dist.)}|\right)\right),
\end{equation}
which straightforwardly leads us to that the QCRBs $\mathrm{Tr}\left\{\mathbf{F}_{\rm Q}^{-1}\left(\bigotimes_{j=0}^{m}\Lambda_{\gamma_j}\left(|\psi_{N,\bm{\phi}}^{\rm (Fock)}\rangle\langle\psi_{N,\bm{\phi}}^{\rm (Fock)}|\right)\right)\right\}$ and $\mathbf{F}_{\rm Q}\left(\bigotimes_{j=0}^{m}\Lambda_{\gamma_j}\left(|\psi_{N,\bm{\phi}}^{\rm (dist.)}\rangle\langle\psi_{N,\bm{\phi}}^{\rm (dist.)}|\right)\right)$ are equal to each other. Since the $N$ distinguishable particles are independent to each other, their QCRB is equal to the QCRB of a single particle multiplied by $1/N$. Also, from that the QCRB of the single particle of equal to $\frac{1}{4}\left\{\sqrt{\frac{m}{1-\gamma_0}}+\sum_{j=1}^{m}\frac{1}{\sqrt{1-\gamma_j}}\right\}^2$, we verify that the QCRB of $N$ distinguishable particles is equal to the SQL $\frac{1}{4N}\left\{\sqrt{\frac{m}{1-\gamma_0}}+\sum_{j=1}^{m}\frac{1}{\sqrt{1-\gamma_j}}\right\}^2$ as derived in Ref.~\cite{m.namkung}. Consequently, the SQL is equivalent to the QCRB of $N$-photon Fock state enterring a multi-mode beam-splitter.
\end{widetext}

\end{document}